\documentclass[twocolumn]{aastex631}
\usepackage{epsfig}
\usepackage{amsmath, amsthm, amssymb}
\usepackage{microtype}
\usepackage{float}
\usepackage{verbatim}
\usepackage{wrapfig}
\usepackage{graphicx}
\usepackage[caption = false]{subfig}
\usepackage{multirow}
\usepackage{hhline}
\usepackage{booktabs}
\usepackage{array}
\usepackage{bm}
\usepackage[normalem]{ulem}
\usepackage{changepage}

\usepackage{rotating}

\shorttitle{A New Huntsman Millisecond Pulsar}
\shortauthors{Strader et al.}

\begin{document}

\title{PSR J1947--1120: A New Huntsman Millisecond Pulsar Binary}

\correspondingauthor{Jay Strader}
\email{straderj@msu.edu}

\author[0000-0002-1468-9668]{Jay Strader}
\affiliation{Center for Data Intensive and Time Domain Astronomy, Department of Physics and Astronomy,\\ Michigan State University, East Lansing, MI 48824, USA}
\author[0000-0002-5297-5278]{Paul S. Ray}
\affiliation{Space Science Division, U.S. Naval Research Laboratory, Washington, DC 20375, USA}
\author[0000-0003-1814-8620]{Ryan Urquhart}
\affiliation{Center for Data Intensive and Time Domain Astronomy, Department of Physics and Astronomy,\\ Michigan State University, East Lansing, MI 48824, USA}
\author[0000-0003-1699-8867]{Samuel J. Swihart}
\affiliation{Science, Systems and Sustainment Division, Institute for Defense Analyses, Alexandria, VA 22305, USA}
\author[0000-0002-8400-3705]{Laura Chomiuk}
\affiliation{Center for Data Intensive and Time Domain Astronomy, Department of Physics and Astronomy,\\ Michigan State University, East Lansing, MI 48824, USA}
\author[0000-0001-8525-3442]{Elias Aydi}
\affiliation{Center for Data Intensive and Time Domain Astronomy, Department of Physics and Astronomy,\\ Michigan State University, East Lansing, MI 48824, USA}
\author[0000-0001-8018-5348]{Eric C. Bellm}
\affiliation{DIRAC Institute, Department of Astronomy, University of Washington, 3910 15th Avenue NE, Seattle, WA 98195, USA}
\author[0000-0002-8532-4025]{Kristen C. Dage}
\affiliation{International Centre for Radio Astronomy Research, Curtin University, GPO Box U1987, Perth, WA 6845, Australia}
\author[0000-0002-2185-1790]{Megan E. DeCesar}
\affiliation{College of Science, George Mason University, Fairfax, VA 22030, USA\\ resident at U.S. Naval Research Laboratory, Washington, DC 20375, USA}
\author[0000-0003-1226-0793]{Julia S. Deneva}
\affiliation{College of Science, George Mason University, Fairfax, VA 22030, USA\\ resident at U.S. Naval Research Laboratory, Washington, DC 20375, USA}
\author[0000-0001-7697-7422]{Maura A. McLaughlin}
\affiliation{Department of Physics and Astronomy, West Virginia University, P.O. Box 6315, Morgantown, WV 26506, USA}
\affiliation{Center for Gravitational Waves and Cosmology, West Virginia University, Chestnut Ridge Research Building, Morgantown, WV 26505, USA}
\author[0009-0004-4418-0645]{Isabella Molina}
\affiliation{Center for Data Intensive and Time Domain Astronomy, Department of Physics and Astronomy,\\ Michigan State University, East Lansing, MI 48824, USA}
\author[0000-0001-8424-2848]{Teresa Panurach}
\affiliation{Center for Data Intensive and Time Domain Astronomy, Department of Physics and Astronomy,\\ Michigan State University, East Lansing, MI 48824, USA}
\affiliation{Norfolk State University, 700 Park Avenue, Norfolk, VA 23504 USA}
\author[0000-0001-5991-6863]{Kirill V. Sokolovsky}
\affiliation{Department of Astronomy, University of Illinois at Urbana-Champaign, 1002 W. Green Street, Urbana, IL 61801, USA}

\begin{abstract}
We present the discovery of PSR J1947--1120, a new huntsman millisecond pulsar with a red giant companion star in a 10.3 d orbit. This pulsar was found via optical, X-ray, and radio follow-up of the previously unassociated $\gamma$-ray source 4FGL J1947.6--1121. PSR J1947--1120 is the second confirmed pulsar in the huntsman class and establishes this as a bona fide subclass of millisecond pulsar. We use {\tt MESA} models to show that huntsman pulsars can be naturally explained as neutron star binaries whose secondaries are currently in the ``red bump" region of the red giant branch, temporarily underfilling their Roche lobes and hence halting mass transfer. Huntsman pulsars offer a new view of the formation of typical millisecond pulsars, allowing novel constraints on the efficiency of mass transfer and recycling at an intermediate stage in the process.

\vspace{10mm}
\end{abstract}

\section{Introduction}

One early discovery from the Fermi Gamma-Ray Space Telescope was that old recycled neutron stars---millisecond pulsars---are efficient GeV $\gamma$-ray emitters \citep{Abdo2009,Abdo2013}, converting a non-negligible fraction of their spindown luminosity into 
$\gamma$-rays. Targeted searches of newly discovered $\gamma$-ray sources in the Galactic field revealed a host of new millisecond pulsar binaries with unexpected demographics: many were in close binaries with hydrogen-rich companions \citep{Ray12,Roberts13}, unlike the longer orbital period white dwarf--millisecond pulsar binaries that previously dominated known systems \citep{Lorimer2008}. It is now clear that these spider pulsar binaries, which gradually erode or even destroy their secondaries, are common among millisecond pulsar binaries, but were rare in pre-Fermi searches because they often have extensive radio eclipses. These spider systems typically show detectable (and sometimes extreme) optical and X-ray variability, and it was quickly realized that this enabled their complementary discovery at a broad range of wavelengths (e.g., \citealt{Cheung2012,Kong2012,Romani2012,Breton13}).

In a targeted optical and X-ray search of the $\gamma$-ray error ellipse of the previously unassociated source 1FGL J1417.7--4407 (hereafter J1417), \citet{Strader15} discovered an X-ray luminous binary, with a heavily stripped red giant in a 5.4 d orbit around an invisible neutron star--mass companion. Owing to its substantial X-ray luminosity ($\gtrsim 10^{33}$ erg s$^{-1}$) and the presence of persistent luminous double-peaked H$\alpha$ emission, they argued that this binary was a mass-transferring system with a subluminous accretion disk, akin to the transitional millisecond pulsar PSR J1023+0038 (e.g., \citealt{Stappers2014}).

Subsequently, \citet{Camilo16} presented the discovery of the millisecond pulsar PSR J1417--4402 as the optically unseen component in this binary. They argued that no disk was present and that a nearer distance would give an X-ray luminosity more consistent with the intrabinary shocks observed in some spiders \citep{Roberts13}. Instead, they suggested the pulsar was in the ``radio ejection" regime \citep{Burderi2002} where the pulsar radiation pressure prevents mass transfer from the inner Lagrangian point of the red giant. \citet{Swihart18} reconciled these views, with a larger distance (and hence high X-ray luminosity) confirmed via a Gaia parallax, but with no evidence for an accretion disk. Instead it appears that J1417 has an unusually luminous intrabinary shock between the red giant wind and pulsar wind, which produces both the X-ray and H$\alpha$ emission. In any case, J1417 was then unique: no other millisecond pulsar binary had a comparable orbital period and evolved H-rich secondary star.

In the variable $\gamma$-ray source 2FGL J0846.0+2820, \citet{Swihart17} discovered a potential doppelganger to J1417, and suggested the moniker ``huntsman" for these systems that are larger than typical spider binaries. 2FGL J0846.0+2820 is spatially coincident with a 8.1 d binary consisting of another partially stripped red giant secondary and an unseen $\sim 2 M_{\odot}$ primary. However, despite a number of pulsar search observations, no millisecond pulsar has yet been confirmed in this binary, so it remains a candidate.

Here we present the discovery of a confirmed second member of the huntsman class: the millisecond PSR J1947--1120, found within the Fermi $\gamma$-ray source 4FGL J1947.6--1121. This pulsar is in a 10.3 d orbit with a heavily stripped red giant secondary. In Sections \ref{sec:obs} and \ref{sec:res} we characterize the binary and in Section \ref{sec:discussion} we discuss the origin of huntsman systems.

\section{Data}
\label{sec:obs}

\subsection{Gamma-rays}
\label{sec:gammarays}
The millisecond pulsar discovery presented in this paper was made via optical, X-ray, and radio follow-up of the $\gamma$-ray source 
4FGL J1947.6--1121 \citep{Fermi22,Ballet2023}. This source is relatively well-localized by Fermi-LAT, with a 95\% error ellipse of 3.3\arcmin$ \times\ 3.0$\arcmin, and was also detected in the previous 2FGL and 3FGL catalogs. In 4FGL-DR4 the source is significantly curved, both for a power law with subexponential cutoff ($3.1\sigma$) and for a LogParabola model ($3.0\sigma$), and shows no significant variability. 4FGL J1947.6--1121 has a 0.1--100 GeV flux of ($3.4\pm0.7) \times 10^{-12}$ erg s$^{-1}$ cm$^{-2}$, corresponding to a luminosity of $L_{\gamma} = (1.2\pm0.2) \times 10^{34}$ erg s$^{-1}$ at a fiducial distance of 5.4 kpc (Section \ref{sec:lc_model}).

\subsection{X-rays}
\label{sec:X-rays}

\subsubsection{Swift}
\label{sec:Swiftobs}

4FGL J1947.6--1121 has 4.2 ksec of \emph{Swift}/XRT observations obtained over seven epochs from 2019 Jul to 2020 May. As catalogued in the Living Swift-XRT Point Source Catalogue \citep{Evans2023} that encompasses these data, a single significant X-ray source is present within the Fermi-LAT error ellipse. This source has J2000 position (R.A., Dec.) of (19:47:37.93, --11:20:27.6) with a 90\% positional uncertainty of 7.1\arcsec. The 1--10 keV count rate is  $2.0\pm1.2 \times 10^{-3}$ cts s$^{-1}$. Given the low count rate, with $\sim 6$ net counts in the 1--10 keV range, no spectral fit or test for variability can be reliably accomplished. For an assumed foreground of $N_H = 9.43 \times 10^{20}$ cm$^{-2}$ \citep{HI4PI} and photon index $\Gamma = 1.8$, this count rate implies an unabsorbed 1--10 keV flux of $1.0\pm0.6 \times 10^{-13}$ erg s$^{-1}$ cm$^{-2}$, corresponding to a luminosity of $L_{X} = 1.3\pm0.8\times 10^{32}$ erg s$^{-1}$ at a distance of 5.4 kpc. 

\subsubsection{XMM}

We obtained an observation of 4FGL J1947.6--1121 with the European Photon Imaging Camera (EPIC) on XMM-Newton from 2024 Apr 12 UT 02:34 to 2024 Apr 13 UT 09:39 (Obs ID 0920540101), with a total exposure time just under 112 ksec. The EPIC MOS1 and MOS2 data were obtained in full frame mode with the thin filter; the EPIC pn data were taken in timing mode.
Here we analyze only the MOS data, deferring analysis of the pn data to a future paper.

We reprocessed the data using standard tasks within the Science Analysis System (SAS; \citealt{Gabriel2004}) version 18.0.0 software package. Intervals of high particle background exposure at the start and end of the observation were filtered out. We applied standard flagging criteria: \verb|FLAG == 0|, \verb|#XMMEA_EM| and \verb|PATTERN <=12|. We used circular source extraction regions of 30$^{\prime\prime}$ and local background regions at least three times larger.

We extracted background-subtracted light curves using the SAS tasks \textit{evselect} and \textit{epiclccorr}. The individual MOS1 and MOS2 light curves were combined using the {\sc ftools} package \textit{lcmath} \citep{Blackburn95}. Background-subtracted spectra were extracted for MOS1 and MOS2 using \textit{xmmselect} before being combined into a single MOS spectrum using \textit{epicspeccombine}. The combined spectrum was grouped to at least 20 counts per bin in order to use Gaussian statistics during spectral fitting, performed in {\sc Xspec} version 12.10.1 \citep{Arnaud96}. 

\subsection{Optical Photometry}
\subsubsection{Gaia}
\label{sec:optcounterparts}
There is only one optical source matching the single \emph{Swift}/XRT X-ray source found within the error ellipse of 4FGL J1947.6--1121, and it is listed as  {Gaia} DR3 4189956032809439488 \citep{GaiaDR3} with a mean $G=16.40$ mag and an ICRS position of (R.A., Dec.) = (19:47:38.238, --11:20:27.21). This source, referred to as J1947 for the remainder of the paper, has a well-measured proper motion of ($\mu_{\alpha}$ cos $\delta$, $\mu_{\delta}$)= ($-0.58\pm0.06$ mas yr$^{-1}$,--$0.99\pm0.05$ mas yr$^{-1}$).
The zeropoint-corrected \citep{Lindegren2021} Gaia DR3 parallax is $\varpi = 0.173\pm0.061$, which implies a distance of $5.7^{+2.0}_{-1.3}$ kpc
for a standard direction-dependent distance prior \citep{BailerJones21}; a uniform prior gives an essentially identical distance. 

As early as Gaia DR2 \citep{GaiaCollaboration2018} we recognized J1947 as a source with a photometric uncertainty that was higher than expected for an isolated non-variable star, suggesting it could be variable \citep{Andrew21,Mowlavi21}, though at that time no X-ray data were available that covered 4FGL J1947.6--1121. In Gaia DR3, 27 epochs of photometry were released for J1947, confirming that it is variable, with $\sigma = 0.04$ mag in $G$.

Our follow-up spectroscopy (Section \ref{sec:spectra}) confirms that this source is indeed the binary companion to the newly-discovered millisecond pulsar.

\subsubsection{Zwicky Transient Factory Photometry}

In addition to the small number of epoch photometry measurements available from Gaia, there are many epochs of photometry from 2018 Apr to 2023 Nov in $g$ and $r$ for J1947 from the Zwicky Transient Factory \citep[ZTF;][]{Bellm2019}. We took the publicly available measurements from ZTF DR21. We removed those flagged as unreliable, and additionally excluded those with large ($> 0.03$ mag) uncertainties on the per-measurement photometric zeropoint. 
Finally, 129 of the $r$ measurements were taken over a 1.4\,h time span on a single night. Given the long orbital period of the binary, no significant variability is expected or observed over this interval; we bin (only) these measurements by a factor of ten, {using the median photometric uncertainty within each bin to represent the bin,} so they do not inappropriately dominate the light curve fitting in Section \ref{sec:lc_model}. This left 468 datapoints in $r$ and 252 in $g$. The epochs of all photometry were converted into Barycentric Modified Julian Dates on the TDB system (henceforth BMJD).

\subsection{Optical Spectroscopy}
\label{sec:SOARdata}
\begin{figure}[t!]
\begin{center}
	\includegraphics[width=\linewidth]{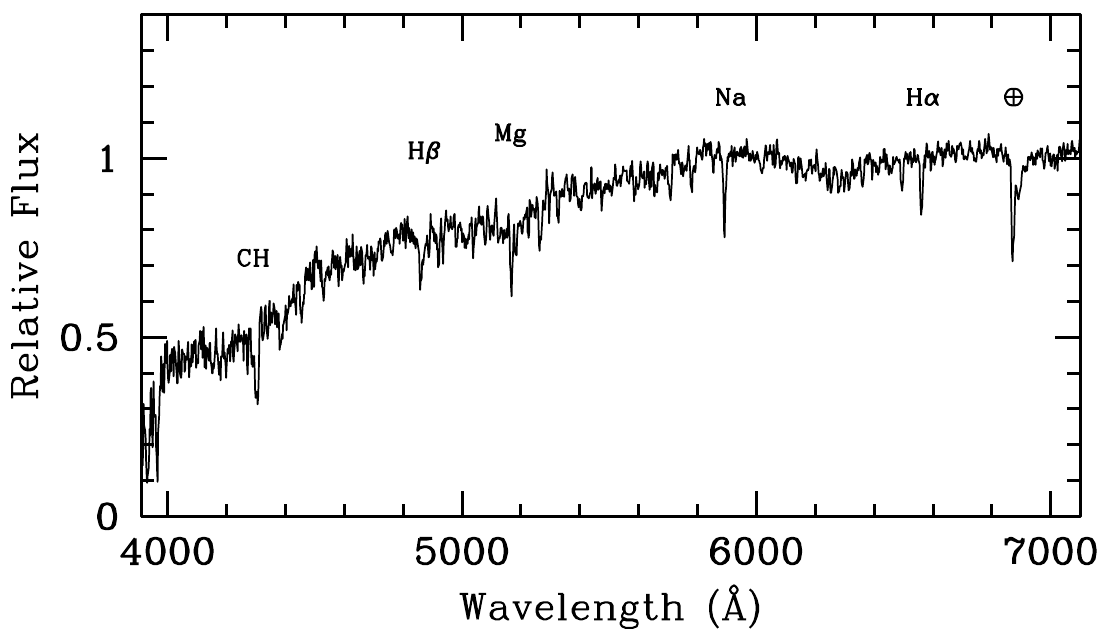}
    \caption{Low-resolution optical spectrum of J1947 from 1 Oct 2021, at $\phi=0.09$. A relative flux calibration has been applied. The spectrum is consistent with a cool K-type star, and the strongest metal and Balmer absorption features along with the telluric Fraunhofer B band are labelled.}
\label{fig:spec_fig}
\end{center}
\end{figure}

We performed spectroscopy of J1947 with the Goodman Spectrograph \citep{Clemens04} on the 4.1-m SOAR telescope from Oct 2021 to Oct 2022. The first spectrum, taken on 1 Oct 2021, used a 400 l mm$^{-1}$ grating and $1.2\arcsec$ longslit, giving a full width at half-maximum (FWHM) resolution of 7.3 \AA\, and showed a K-type spectrum dominated by metal lines (Figure \ref{fig:spec_fig}). We then began spectroscopic monitoring of the source to measure radial velocities, obtaining an additional 39 usable spectra on 19 different nights. These monitoring spectra all used the red camera and a 2100 l mm$^{-1}$ grating, covering the wavelength range $\sim 6100$--6650 \AA\ at a FWHM resolution of either 1.0 \AA\ (for the $1.2\arcsec$ longslit) or 0.85 \AA\ (for the $1.0\arcsec$ longslit). The exposure time per spectrum was 20 min. The spectra were reduced and optimally extracted using standard methods in \textsc{IRAF} \citep{Tody86,Tody93,Fitzpatrick24}.

\subsection{Radio Pulsar Search Data}

Following the detection of periodic optical photometric and radial velocity variations from the candidate, we obtained a series of pulsar search and timing observations with the 100-m Robert C. Byrd Green Bank Telescope over 14 epochs from 2021 Dec 22 to 2023 Jun 9. The first two observing blocks were 45 min in length, and the remainder were 120 min. All observations were made with the  PF1 receiver at 820 MHz and the VEGAS backend, with 200 MHz of bandwidth over 4096 channels and an integration time of 81.92 $\mu$s.

\section{Results}
\label{sec:res}

\subsection{Pulsar Detection and Timing}
\label{sec:timing}

\begin{figure}[t!]
\begin{center}
    \includegraphics[width=\linewidth]{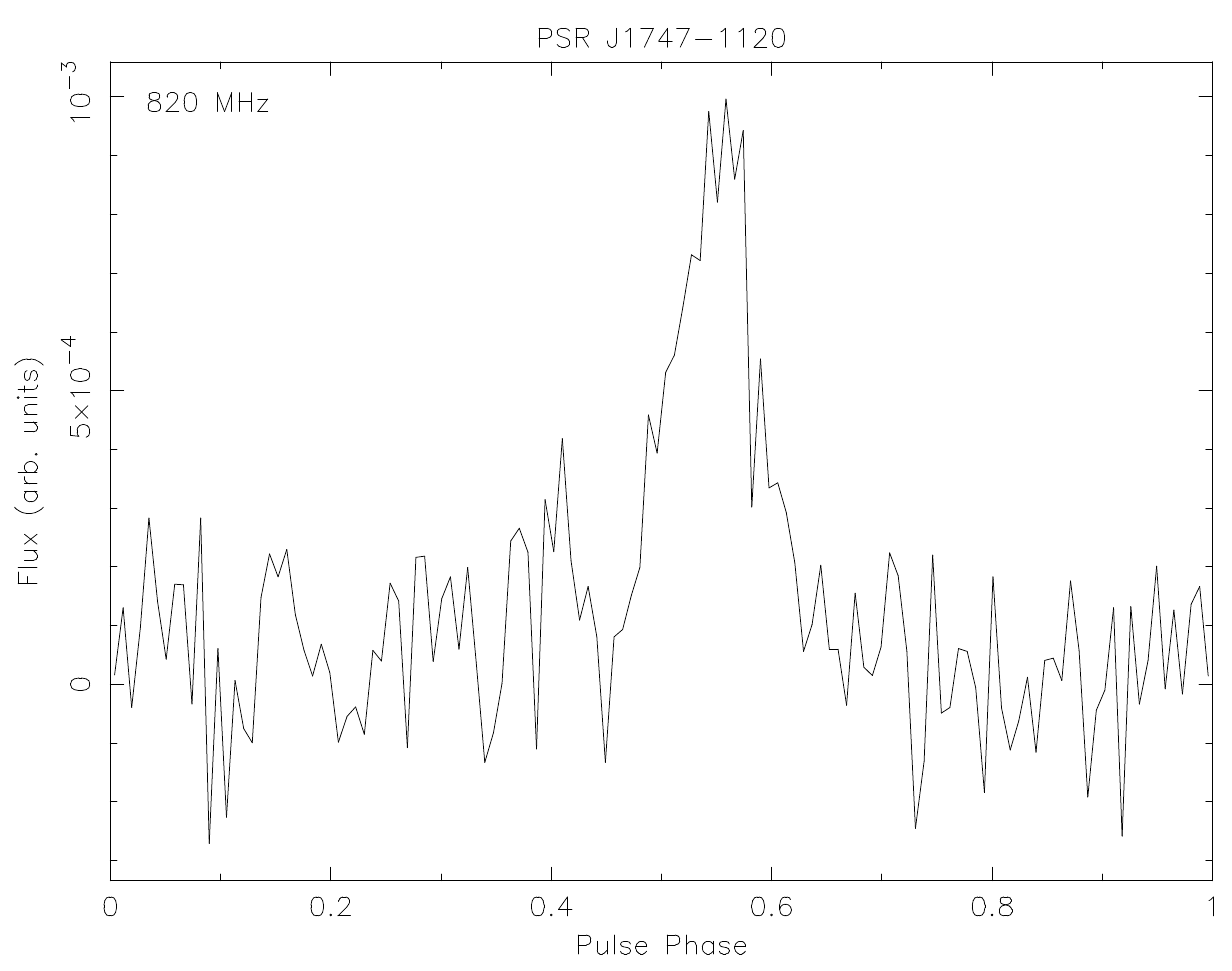}
    \caption{Dedispersed mean pulse profile for PSR J1947--1120 from the GBT discovery observation on 2021 Dec 22}
    \label{fig:j1947_psr}
\end{center}
\end{figure}

Our initial data analysis was done in \textsc{PRESTO} v4.0 \citep{Ransom11}. After removal of RFI, we performed an acceleration search for periodic signals.
A millisecond pulsar was detected in the first observing epoch (2021 Dec 22) at high significance (Figure~\ref{fig:j1947_psr}), with a spin period of 2.24 ms and a dispersion measure of 50.85 pc cm$^{-3}$. It was confirmed in a subsequent observation on 2022 Feb 7. The remainder of the observations were made to time the pulsar.

The pulsar was detected in 11 of the 14 observations. Fixing the position and proper motion to the Gaia values, and using initial values from the optical spectroscopy (Section \ref{sec:spectra}) for the orbital period and time of the ascending node, we attempted to find a phase-connected timing solution to the TOAs using \textsc{APTB} \citep{Taylor2024}, which in turn makes use of \textsc{PINT} \citep{Luo2021}.

We found a number of fits that were reasonably good, with $\chi^2$/d.o.f. $\sim 1.5$--1.6 without any scaling of the TOA uncertainties, but were unable to find a fully phase-connected solution. We also considered fits that had JUMPs around each observation. 
The best JUMPed fit, which assumed a circular orbit and a typical millisecond pulsar spindown frequency of $10^{-15}$ s$^{-2}$, had a spin period 2.240104464(6) ms, orbital period 10.264188(7) d, BMJD epoch of the ascending node of the pulsar 59569.2919(3), and $a$ sin $i = 6.835(5)$ lt-s. 

The best solutions without JUMPs fell roughly into two groups, one with high $\dot{P}$ and hence large inferred minimum spindown luminosities ($\sim 10^{36}$--$10^{37}$ erg s$^{-1}$, assuming a neutron star mass of 1.4 $M_{\odot}$) and pulsar magnetic field ($\sim 10^9$ G) and the other with a lower $\dot{P}$, spindown luminosity ($\sim 3 \times 10^{34}$  erg s$^{-1}$) and magnetic field ($\sim 10^8$ G). Given the properties of other millisecond pulsars observable as Fermi $\gamma$-ray sources \citep{Smith2023}, the latter family of solutions appears much more likely. The parameters of the best-fit preliminary timing solution from this family are: spin period 2.240104451(2) ms, orbital period 10.26421(2) d, BMJD epoch of the ascending node of the pulsar 59569.2915(6), pulsar projected semi-major axis $a$ sin $i = 6.835(3)$ lt-s, and eccentricity 0.0004(2). The uncertainties given represent typical variations among sets of low $\chi^2$/d.o.f. solutions from different groups of solutions, and should be taken as broadly indicative rather than precise uncertainties. We emphasize that these values are not from a phase-connected solution and that additional timing observations are needed.

\subsubsection{Pulsar Eclipses and Non-Detections}

The pulsar was not detected in three of the 14 epochs: 2021 Dec 23, 2022 Apr 5, and 2022 Nov 1. The first two of these occur very close to conjunction ($\phi = 0.249$ and 0.246) with the red giant in front of the pulsar. This is exactly when pulsar eclipses due to scattering or absorption from extended ionized material from the companion is most likely. {Note that the uncertainty in the orbital period corresponds to an uncertainty of only $4 \times 10^{-5}$ in orbital phase over the 534 d timespan of the timing observations, negligible for these comparisons.}
The phase of the 2022 Nov 1 non-detection is quite different ($\phi = 0.67$), though eclipses have been observed at a wide range of phases in pulsars with close hydrogen-rich companions  (e.g., \citealt{Corongiu2021}). Scintillation is also a possibility. In any case, the eclipses for J1947 are much less extensive than for the other huntsman pulsar J1417, which is eclipsed the majority of the time \citep{Camilo16}. This suggests a weaker intrabinary shock or different geometry for J1947 compared to J1417.

\subsubsection{Dispersion Measure Distance}

The pulsar dispersion measure of 50.85 pc cm$^{-3}$ gives a distance of 3.1 kpc using the YMW16 model \citep{Yao17} or 1.9 kpc using the NE2001 model \citep{Cordes02}. These values are lower than those derived from the Gaia parallax (Section \ref{sec:optcounterparts}) or light curve fitting (Section \ref{sec:lc_model}), which in turn agree well with each other. Previous works have found that for binary pulsars with detectable optical counterparts, including spider systems, that parallax and light curve distances are more accurate than dispersion measure distances \citep{Jennings2018,Koljonen2023}. Hence we do not use the dispersion measure-based distance for the remainder of the paper, instead using the light curve fitting distance of $5.4\pm0.3$ kpc, which is consistent with the  Gaia parallax distance of $5.7^{+2.0}_{-1.3}$ kpc.

\subsection{Optical Spectroscopy}
\label{sec:spectra}

The SOAR spectra all look very similar: a K star with a forest of strong absorption lines. The H$\alpha$ region is included in all the spectra, but there is no evidence for H$\alpha$ emission in any of them.

We fit absorption-line radial velocities using {\tt RVSpecFit} \citep{Koposov2011,Koposov2019}, which performs full spectral fitting over a grid of PHOENIX \citep{Allard2016} templates convolved to the observed resolution. These are listed in Table~\ref{table:rv}.

\begin{deluxetable}{crr}[!t]
\label{table:rv}
\tablecaption{Optical Radial Velocities}
\tablehead{
\colhead{BMJD\tablenotemark{a}} & \colhead{radial velocity} & \colhead{unc.} \\
(d) & (km s$^{-1}$) & (km s$^{-1}$)
}
\startdata
59488.1704063 & --86.9 & 1.6 \\
59488.1844094 & --86.3 & 1.8 \\
59494.1389264 & 14.3 &  1.5 \\
59494.1531714 & 11.1 &  1.5 \\
59518.0280666 & --98.8 & 1.4 \\
59518.0421852 & --100.6 & 1.4 \\
59522.0276554 & 43.1 &  1.4 \\
59522.0418555 & 42.4 &  1.4 \\
59532.0187035 & 36.8 &  1.5 \\
59532.0330559 & 36.7 &  1.5 \\
59533.0165177 & 58.0 &  1.4 \\
59533.0308680 & 55.7 &  1.4 \\
59667.4026503 & 55.4 &  1.4 \\
59680.3870426 & --57.7 & 1.4 \\
59680.4012704 & --55.0 & 1.4 \\
59684.3044315 & --41.8 & 1.4 \\
59684.3184784 & --40.3 & 1.4 \\
59700.3973118 & --31.0 & 1.5 \\
59700.4113816 & --32.1 & 1.5 \\
59724.3006529 & --85.3 & 1.4 \\
59724.3148576 & --81.4 & 1.4 \\
59740.2664580 & 25.4 &  1.4 \\
59740.2804646 & 24.6 &  1.4 \\
59807.2386002 & --53.4 & 1.4 \\
59807.2526543 & --52.5 & 1.4 \\
59807.2751650 & --52.8 & 1.4 \\
59807.2892199 & --52.1 & 1.4 \\
59816.2045466 & --92.7 & 1.4 \\
59816.2186680 & --93.1 & 1.4 \\
59840.1561963 & 42.2 &  1.4 \\
59840.1738609 & 40.2 &  1.4 \\
59850.1533645 & 33.2 &  1.5 \\
59850.1675357 & 32.9 &  1.5 \\
59871.0926497 & 46.5 &  1.4 \\
59871.1068424 & 44.9 &  1.4 \\
59513.0751315 & 60.1 &  1.4 \\
59513.0898544 & 60.7 &  1.5 \\
59882.0578174 & 60.4 &  1.4 \\
59882.0718205 & 58.7 &  1.4 \\
\enddata
\tablenotetext{a}{Barycentric Modified Julian Date}
\end{deluxetable}

Using {\tt TheJoker} \citep{Price-Whelan2017}, we fit a circular Keplerian model to the optical radial velocities alone. The best-fitting parameters are period $P=10.26506\pm0.00063$ d, secondary velocity semi-amplitude $K_2 = 80.0\pm0.3$ km s$^{-1}$, systemic velocity $-19.7\pm0.3$ km s$^{-1}$, and BMJD epoch of the ascending node of the neutron star $T_0 = 59569.280\pm0.011$ d, with $1\sigma$ Gaussian uncertainties listed. This is an exceptionally good fit, with an r.m.s. of only 1.4 km s$^{-1}$ and a $\chi^2/$d.o.f. = 35.5/35. The mass function of the pulsar is $f(M) = 0.543\pm0.007 M_{\odot}$. The orbital eccentricity implied by the preliminary pulsar timing in Section \ref{sec:timing} is $< 10^{-3}$, so a circular fit is adequate.

\begin{figure}[t!]
\begin{center}
    \includegraphics[width=\linewidth]{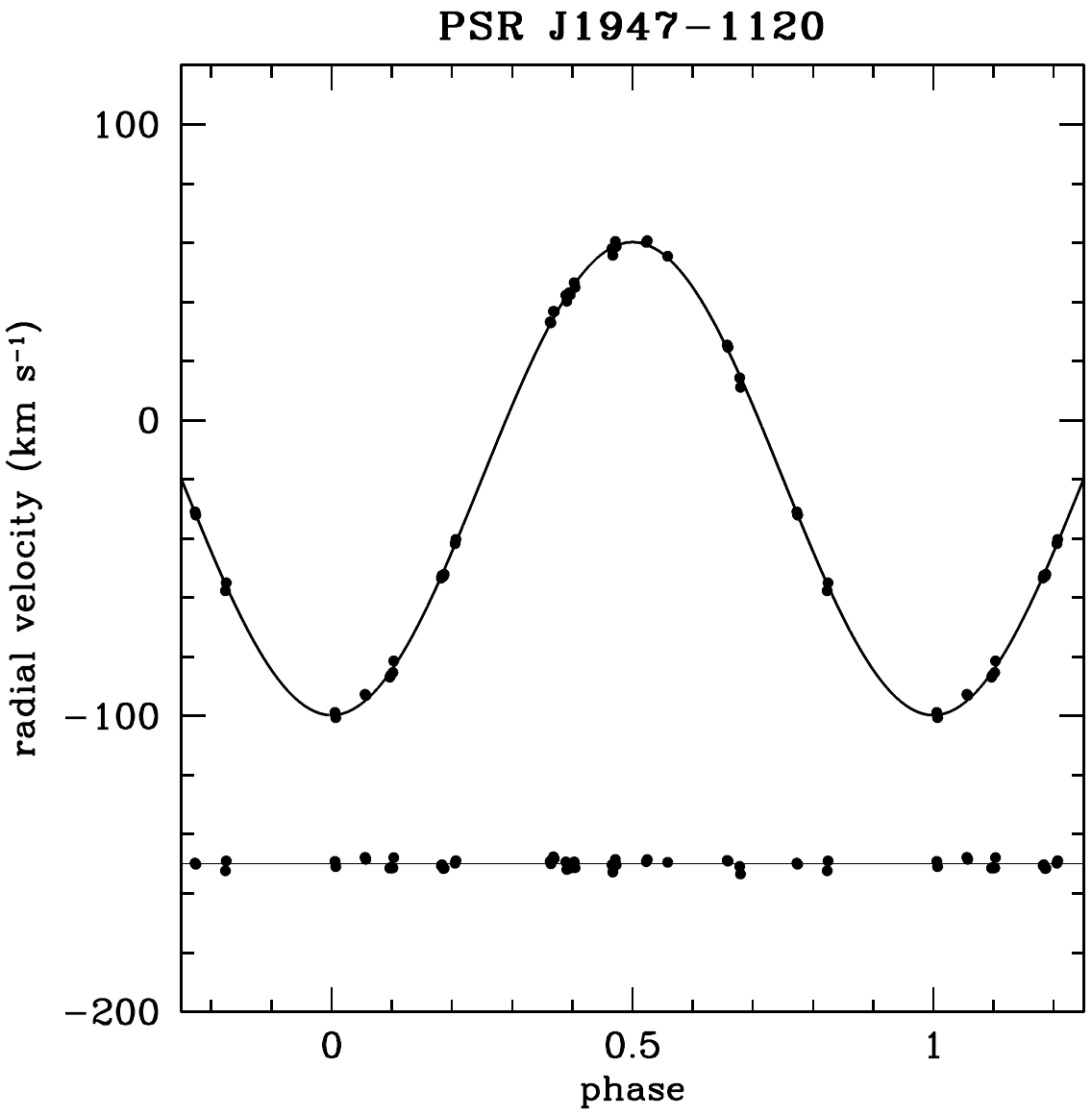}
    \caption{Circular Keplerian fit to the SOAR absorption-line radial velocities of J1947. The fit residuals are plotted offset below the model.}
    \label{fig:j1947_rv}
\end{center}
\end{figure}

This orbital period and epoch of the ascending node are consistent with, but of lower precision than, that available from the pulsar timing observations even 
without a phase-connected timing solution.  If we fix these to the values from the most plausible timing solution from Section \ref{sec:timing}, it has an essentially identical r.m.s. scatter, $K_2$, systemic velocity, and $\chi^2/$d.o.f. = 37.5/37. We show this fit in Figure~\ref{fig:j1947_rv}. The results from the optical spectroscopy and pulsar timing are in complete agreement, confirming this is indeed the binary companion to the pulsar.

Owing to the relatively long period of the binary and the high luminosity of the red giant, irradiation is expected to minimally affect the secondary. Hence it is a reasonable assumption that the measured $K_2$ of the secondary reflects the motion of its center of mass. In this case, the combination of pulsar $a$ sin $i$, orbital period, and secondary $K_2$ directly gives the mass ratio $q = M_2/M_1 = 0.182(1)$.

\subsection{Light Curve Modeling}
\label{sec:lc_model}

\begin{figure}[t!]
\begin{center}
    \includegraphics[width=\linewidth]{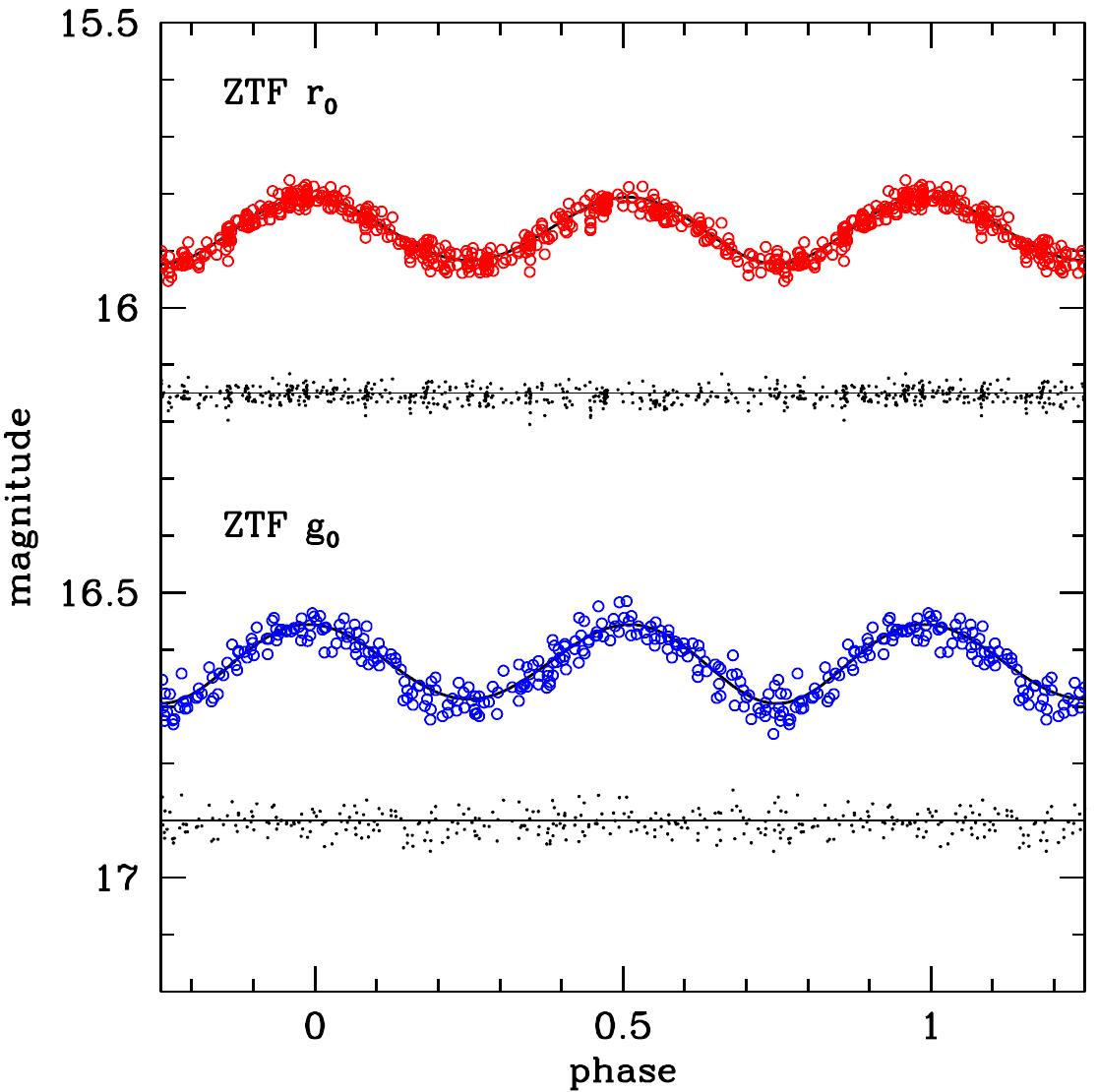}
    \caption{ZTF $g_0$ and $r_0$ photometry of J1947 with the best-fit ellipsoidal model overplotted. The fit residuals are shown under each light curve.}
    \label{fig:j1947_lc}
\end{center}
\end{figure}

The phased ZTF light curve is shown in  Figure~\ref{fig:j1947_lc}, which shows low-amplitude ellipsoidal modulations. As noted above, it is reasonable to neglect irradiation in modeling this system.

We modeled the light curve using {\tt PHOEBE} version 2.4.14 \citep{Prsa2005,Conroy2020}. {The orbital period, epoch of the ascending node, and neutron star $a$ sin $i$, were fixed to the values determined from the pulsar timing, and the binary mass ratio was fixed to the value determined jointly from timing and optical spectroscopy. While the latter value of $q=0.182\pm0.001$ is less precise than the quantities solely dependent on pulsar timing, it corresponds to an uncertainty in the secondary Roche lobe radius of about $0.2\%$, negligible in the context of the other uncertainties in the light curve fitting.} {The only prior used in the fitting was $E(g-r) = 0.17\pm0.02$ for the foreground reddening \citep{Green2019}.} This gives a median $r_0 = 15.86$ mag and $(g-r)_0 = 0.76$ mag. 

The parameters fit were the mass of the neutron star (restricted to 1.4--2.1 $M_{\odot}$), the effective temperature of the secondary, the Roche lobe filling factor of the secondary, and the distance. Given the known parameters, the neutron star mass uniquely determines the binary inclination, which must lie in the range $i \sim 47$--$55^{\circ}$. We assume solar metallicity {and model atmospheres from \citet{Castelli2003}}.

We find best-fitting values of $T_{\rm eff} = 4534\pm41$ K, filling factor $0.87\pm0.02$, and distance $5.36\pm0.37$ kpc, {with a goodness of fit $\chi^2$/d.o.f = 1.19 (854/712)}. This model is shown in Figure~\ref{fig:j1947_lc}.
Unfortunately, the mass of the neutron star is essentially unconstrained by the light curves in the context of the other constraints. While it sets the physical size scale for the system, with a more massive neutron star giving a larger Roche lobe for the secondary, the filling factor and distance strongly covary to produce an essentially identical light curve, with a minor contribution from the associated inclination change. The $T_{\rm eff}$ of the secondary is mostly determined by the color and hence is insensitive to this covariance. The distance inferred from this fitting is fully consistent with, but notionally more precise than, the Gaia parallax distance of $5.7^{+2.0}_{-1.3}$ kpc (Section \ref{sec:optcounterparts}). {We note that the listed uncertainties in the inferred parameters from the light curve modeling do not capture all systematic uncertainties, including the use of a single set of model atmospheres as well as the assumed ZTF filter curves and zeropoints.}

To check the effects of a modest metallicity change on the results, we repeated the fitting for [Fe/H] = --0.5, finding---as would be expected---a slightly lower $T_{\rm eff} = 4472\pm40$ K and a closer distance of $5.05\pm0.32$ kpc, but no meaningful change to the inferred filling factor.

The inferred bolometric luminosity of the red giant secondary (for solar metallicity) is $10.9\pm2.1 L_{\odot}$, with the uncertainty dominated by the uncertainty in the light curve-derived distance. For red giants, the bolometric luminosity is determined by the core mass alone. Using the relation from \citet{Boothroyd1988} appropriate for this luminosity range, we find a core mass of $M_c = 0.19\pm0.01 M_{\odot}$ (this is also for solar metallicity, but the relation is only weakly metallicity-sensitive). Compared to the total (core+envelope) red giant mass of $\sim 0.25$--$0.4 M_{\odot}$ implied by a neutron star mass range of 1.4--2.1 $M_{\odot}$, it is clear the red giant has been heavily stripped, leaving an envelope of only $\sim 0.06$--$0.2 M_{\odot}$. This is comparable to the stripping
inferred for the other confirmed huntsman millisecond pulsar, J1417 \citep{Strader15,Camilo16,Swihart18}.

\subsection{X-ray Spectrum and Light Curve}
\label{sec:xray_results}

We fit an absorbed power-law to the XMM MOS X-ray spectrum, finding $N_H = 2\pm1 \times 10^{21}$ cm$^{-2}$ and a photon index of $\Gamma = 1.9\pm0.3$. The fit is good, with a $\chi^2$/d.o.f of 29.7/40. Adding a thermal component does not meaningfully improve the spectral fit. For the power-law fit, the 1--10 keV unabsorbed flux is $4.0\pm0.6 \times 10^{-14}$ erg s$^{-1}$ cm$^{-2}$. This corresponds to a luminosity $L_X = 5.2\pm0.8 \times 10^{31}$ erg s$^{-1}$, consistent with that inferred from the Swift/XRT data (Section \ref{sec:Swiftobs}) within the large uncertainties of the latter.

An $L_X \sim 5 \times 10^{31}$ erg s$^{-1}$ is broadly consistent with that observed for redback millisecond pulsars (e.g., \citealt{Linares2014,Roberts2015,Hui2019,Strader19,Urquhart2020,Swihart2022}) and modeled by emission from an intrabinary shock (e.g., \citealt{vanderMerwe2020}). However, this $L_X$
is much lower than for the huntsman J1417, which has $L_X \sim 10^{33}$ erg s$^{-1}$, a harder photon index of $\Gamma = 1.4\pm0.1$, and broad H$\alpha$ emission likely from the shock \citep{Swihart18}. The lower X-ray luminosity and softer X-ray spectrum for J1947, as well as the lack of H$\alpha$ emission, indicate a much weaker shock than in J1417.

\begin{figure}[t!]
\begin{center}
    \includegraphics[width=\linewidth]{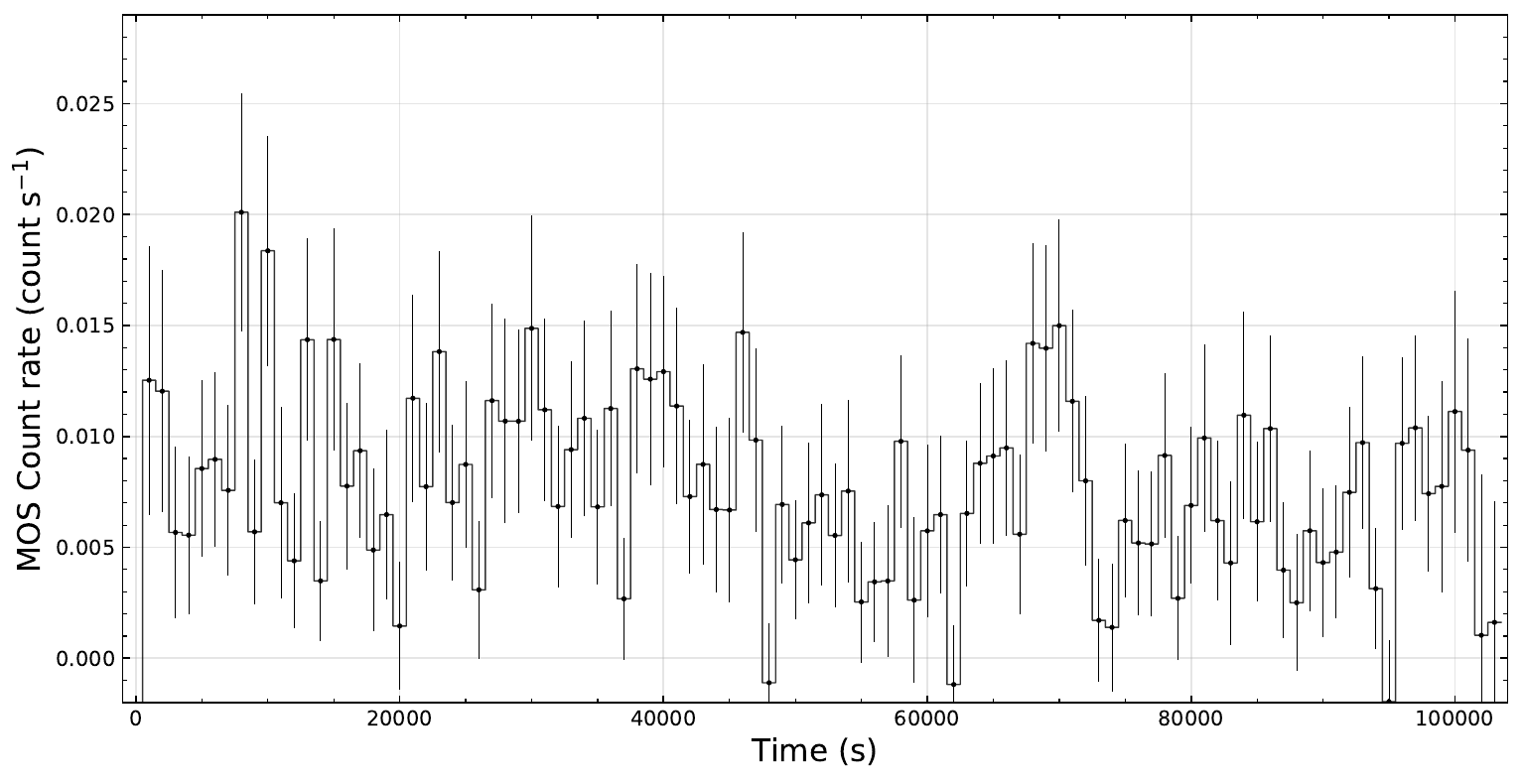}
    \caption{XMM MOS light curve of J1947, from phases $\phi \sim 0.11$ to 0.23 (BMJD 60412.1251 to 60413.3288). There is perhaps minor short timescale variability, but no larger overall changes, including near conjunction toward the end of the light curve.}
    \label{fig:xray_lc}
\end{center}
\end{figure}

Figure~\ref{fig:xray_lc} shows the X-ray light curve of the MOS data, with no apparent trends during the timescale of the data.
The XMM data cover phases $\phi \sim 0.11$--0.23, where $\phi=0.25$ is conjunction with the secondary in front of the pulsar. If the pulsar shock was oriented toward the companion, then the X-ray shock luminosity would be broadly expected to increase around $\phi=0.25$, while if it were oriented toward the pulsar, then the increase would occur around $\phi=0.75$ (modulo eclipse effects), as has been observed for some redbacks, (e.g., \citealt{AlNoori2018}). A peak around $\phi = 0.25$ is ruled out by our data; since we have no data around $\phi=0.75$, we cannot assess that phase for J1947.  It is also possible that the intermediate inclination of the binary partially mutes the Doppler boosting of the shock.

It is not immediately clear why the intrabinary shock is weaker for J1947 compared to J1417. The longer orbital period of J1947 has both certain and potential effects on the shock: the most straightforward is simply that the pulsar and red giant have a larger physical separation. As another effect, it is plausible that the lower rotation rate of the tidally-locked red giant in J1947 compared to J1417 has resulted in a weaker dynamo and hence both a weaker wind from the secondary and magnetic field at the intrabinary shock. Another possibility is that some property of the pulsar wind also differs between the pulsars, though their 
$\gamma$-ray luminosities are relatively similar.

\section{Discussion: How to Make a Huntsman}
\label{sec:discussion}

There is now a definite subclass of two confirmed millisecond pulsar--stripped red giant binaries, with an additional candidate system that has at least some comparable properties but has no detected pulsar. The similarities among these binaries, especially for the two confirmed systems, is notable (see Table \ref{tab:hunts}). The millisecond pulsars are both fully recycled, with spins of 2.7 and 2.2 ms for J1417 and J1947, respectively. They have nearly identical binary mass ratios and heavily stripped red giant secondaries around $\sim 0.3 M_{\odot}$. These red giants both underfill their Roche lobes by $\sim 10$--15\%. Finally, their binary periods and luminosities are in a relatively narrow range (5.4 d and 5 $L_{\odot}$ for J1417; 10.3 d and 11 $L_{\odot}$ for J1947): narrow given the context that millisecond pulsar--white dwarf binaries span an orbital period range orders of magnitude larger.

\begin{deluxetable}{lrr}[]
\label{tab:hunts}
\tablecaption{Confirmed Huntsman Millisecond Pulsars}
\tablehead{
\colhead{Property} & \colhead{J1417} & \colhead{J1947} }
\startdata
Orbital Period (d) & 5.374 & 10.265  \\
$M_2$ ($M_{\odot}$) & $0.28^{+0.07}_{-0.03}$ & $0.32\pm0.03$  \\
$R_2$ ($R_{\odot}$) & $3.7\pm0.3$ & $5.4\pm0.3$  \\
$L_2$\tablenotemark{a}  ($L_{\odot}$) & $5.2\pm1.0$ & $10.9\pm2.1$ \\
$M_{\rm 2,core}$ ($M_{\odot}$) & $0.16\pm0.01$ & $0.19\pm0.01$ \\
$T_2$\tablenotemark{b}  (K) & $4560^{+460}_{-336}$ & $4534\pm41$  \\
$f_2$\tablenotemark{c}  & $0.83^{+0.05}_{-0.07}$ & $0.87\pm0.02$ \\
Mass Ratio ($M_2/M_1$) & $0.171\pm0.002$ & $0.182\pm0.001$  \\
LC distance\tablenotemark{d} (kpc) & $3.1\pm0.6$ & $5.4\pm0.4$ \\
Gaia distance (kpc) & $4.2^{+1.0}_{-0.7}$ & $5.7^{+2.0}_{-1.3}$ \\
$P_{\rm spin}$ (ms) & 2.664 &  2.240 \\
$a$ sin $i$\tablenotemark{e}  (lt-s) & $4.876\pm0.009$ & $6.835\pm0.003$ \\
$L_{X}$\tablenotemark{f}  ($10^{32}$ erg s$^{-1}$)  & $10.0^{+0.6}_{-0.4}$  & $0.52\pm0.08$ \\
$L_{\gamma}$\tablenotemark{g} ($10^{34}$ erg s$^{-1}$)& $0.9\pm0.1$  & $1.2\pm0.2$\\ 
\enddata
\tablenotetext{a}{Bolometric luminosity of secondary.}
\tablenotetext{b}{Effective temperature of secondary.}
\tablenotetext{c}{Roche lobe filling factor of secondary.}
\tablenotetext{d}{Best-fitting distance from light curve modeling.}
\tablenotetext{e}{Projected semi-major axis of the pulsar.}
\tablenotetext{f}{1--10 keV unabsorbed X-ray luminosity. Uncertainties do not include distance uncertainty.}
\tablenotetext{g}{0.1--100 GeV luminosity from 4FGL-DR4. Uncertainties do not include distance uncertainty.}
\tablecomments{Unless otherwise stated, values for J1417 are from \citet{Camilo16} or \citet{Swihart18}; those for J1947 are from the present work. All luminosities assume the light curve distance.}
\end{deluxetable}

Neutron star--main sequence binaries with low-mass secondaries ($M_2 \lesssim 1.5 M_{\odot}$) and initial orbital periods longer than
the ``bifurcation period" of $\sim 2$--3 d fill their Roche lobes on the red giant branch and evolve to longer periods as mass transfer occurs on the shell-burning nuclear evolution timescale \citep{Pylyser1988,Tauris99,Podsiadlowski2002}. These systems should appear as low-mass X-ray binaries for the duration of mass transfer and hence not be detectable as radio pulsars.

As discussed in the Introduction, \citet{Camilo16} suggested that J1417 could instead be in the radio ejection regime, where the pulsar radiation could directly prevent mass transfer from the secondary. While this explanation is plausible for a single system, it would not seem to predict that such systems would fall into a narrow range of orbital period, as is the case so far for huntsman binaries. This motivates the consideration of other models.

\subsection{The Red Bump}

As single low-mass stars ascend the red giant branch, there is an apparent caesura in their evolution: the red bump. This occurs as the evolving H-burning shell encounters a discontinuity in the H abundance left behind at the maximum extent of the penetration of the convective envelope \citep{Thomas1967,Iben1968,Sweigart1978,Christensen-Dalsgaard2015}. The star temporarily becomes slightly less luminous and shrinks before eventually resuming its ascent up the giant branch. Stars appear to pile up, causing a bump in the red giant luminosity function that has been observed in many star clusters (e.g., \citealt{King1985,FusiPecci1990}).

A number of authors have shown that in the context of a low-mass X-ray binary, mass transfer should temporarily halt while the secondary traverses the red bump phase, since it has contracted and no longer fills its Roche lobe. This phase is explicitly noted by \citet{Tauris99} and \citet{Podsiadlowski2002}, and even earlier by \citet{Kippenhahn1967} in the context of a close binary without a compact object.

Here we show that for plausible initial conditions, the predicted properties of these red bump binaries closely match those of both confirmed huntsman millisecond pulsar systems and are a natural explanation for their origin.

\subsection{{\tt MESA} Modeling}

We modeled binaries using {\tt MESA} \citep{Jermyn2023} release r23.05.1, assuming a neutron star primary with an initial mass of 1.4 $M_{\odot}$ and a solar metallicity ZAMS secondary star of 1.0 $M_{\odot}$. The \citet{Kolb1990} mass loss scheme was used. {Following the variable descriptions from \citet{Tauris06}, we assumed non-conservative mass transfer with $\alpha = 0.2$ and $\beta = 0.5$, which are the fractions of mass loss from the vicinities of the donor and accretor, respectively, and no mass loss from a circumbinary toroid ($\gamma$). This gave a mass transfer efficiency of $1-\alpha-\beta-\gamma = 0.3$.} We also adopted the standard magnetic braking prescription from \citet{Rappaport1983} with an index of $\gamma = 3$.

For a secondary with an initial mass of $1.0 M_{\odot}$ and the assumptions above, the initial bifurcation period is around 2.6 d. Above this orbital period the secondary fills its Roche Lobe and begins mass transfer past the main sequence turnoff, either as a subgiant (for a narrow range of orbital periods) or as a red giant. 

All of these donors show the red bump behavior at some luminosity on the red giant branch. For those that have already filled their Roche Lobe and initiated mass transfer before they reach the red bump, the models do indeed show a pause in mass transfer, as expected based on previous work. Since the huntsman millisecond pulsars are observed to be fully recycled, with spins $< 3$ ms, we assume that the neutron star needs to have accreted at least $\sim 0.1 M_{\odot}$ to reach these spins \citep{Tauris2012} before the secondary reaches the red bump. Only models with initial orbital periods $< 7$ d show at least $0.1 M_{\odot}$ of accretion onto the neutron star. For initial orbital periods $\gtrsim 10$ d, the secondary has not yet filled its Roche lobe before the red bump region, and thus never pauses its mass transfer due to this effect.

For the plausible huntsman progenitors---those with initial orbital periods in the range 2.6 to 7.0 d---the predicted orbital periods during the huntsman phase range from about 4.5 to 14.5 d. There is a near-monotonic relation between the initial period and the properties of the system at the red bump. The longer initial period systems have higher luminosities, less stripped secondaries, and longer orbital periods.

\begin{figure}[t!]
\begin{center}
    \includegraphics[width=\linewidth]{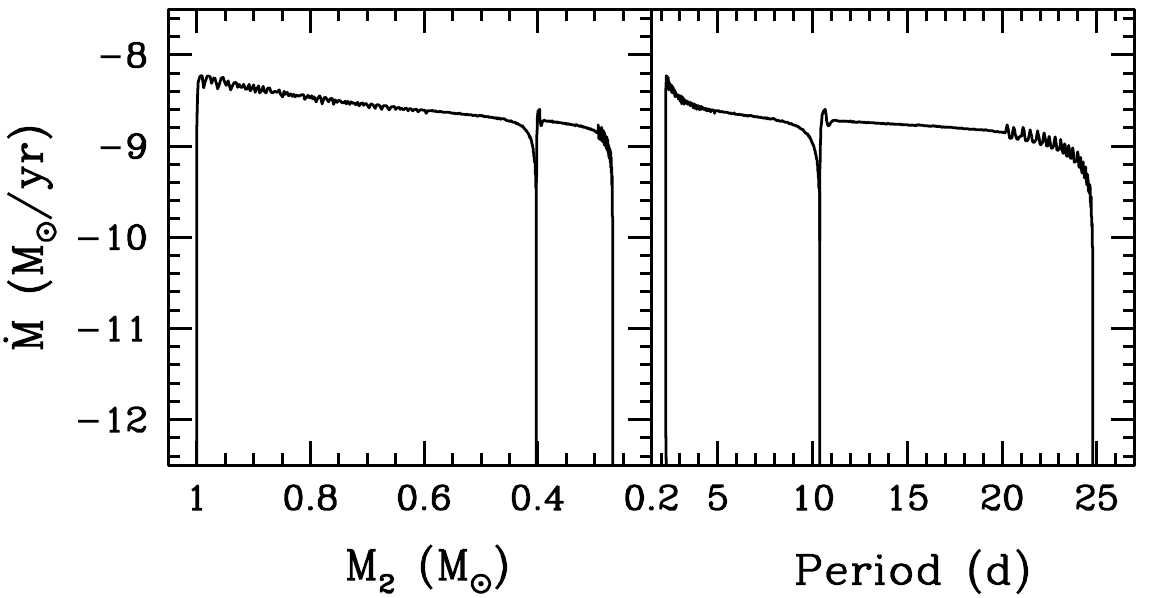}
    \caption{Evolution of the mass-transfer rate to a $1.4 M_{\odot}$ neutron star from the secondary ($\dot{M}$) as a function of secondary mass ($M_2$; left panel) and orbital period (right panel), for a model with initial 
    $M_2 = 1.0 M_{\odot}$ and orbital period of 3.3 d. The temporary cessation of mass transfer during the red bump, at an orbital period of 10.3 d and when $M_2 \sim 0.4 M_{\odot}$, is a reasonable match to the properties of the huntsman millisecond pulsar J1947. This model concludes its evolution as a pulsar--He white dwarf binary with an orbital period of 24.8 d.}
    \label{fig:j1947_mesa}
\end{center}
\end{figure}

In Figure~\ref{fig:j1947_mesa} we show a model in this initial orbital period range that appears to be a close match to the properties of J1947.
{The inlists used to produce this model are publicly available on Zenodo: \dataset[doi:10.5281/zenodo.14518297]{https://doi.org/10.5281/zenodo.14518297}.} From an initial orbital period of 3.3 d, it evolves onto the red giant branch and first fills its Roche lobe after 11.8 Gyr. About 230 Myr later, the secondary reaches the red bump and detaches, causing mass transfer to cease. At this point the orbital period is 10.4 d. The luminosity of the red giant is $\sim 13 L_{\odot}$ and it has already been stripped to a mass of $0.40 M_{\odot}$. The neutron star has accreted $0.17 M_{\odot}$, recycling it to a millisecond pulsar that is visible as a radio pulsar during the $\sim 31$ Myr duration of the red bump phase.

The properties of J1417, which has a lower red giant luminosity of $\sim 5 L_{\odot}$ and shorter current orbital period of 5.4 d (Table \ref{tab:hunts}), are well-matched by a model with a shorter initial orbital period of 2.68 d, just above the bifurcation period. We note that a low-mass X-ray binary model published in \citet{Tauris99}, with a $1.0 M_{\odot}$ donor star and a $1.4 M_{\odot}$ neutron star accretor (see their figure~2), also reaches the red bump with properties comparable to J1417, in their model starting from a slightly longer orbital period of 3.0 d.

This comparison helps to show that all of the specific orbital period values and evolved masses listed above depend on the detailed assumptions used in the models, especially the efficiency of mass transfer and to a lesser degree magnetic braking.
A more comprehensive evaluation of model predictions would be valuable. We have also not explored larger donor masses: more massive secondaries, up to $\lesssim 1.5 M_{\odot}$, should also produce huntsman binaries, though the duration of this phase will decrease at higher masses. 

Nonetheless, these sample calculations demonstrate that the existence of huntsman millisecond pulsars is a straightforward prediction of stellar and binary evolution. This is supported by the close match between the model predictions and observations.

\section{Conclusions and Future Work}

We have presented the discovery and characterization of a new millisecond pulsar binary. It is the second confirmed in the huntsman class, which have partially stripped red giant donors in $\mathcal{O} \sim 10$ d orbits. We have also shown that the existence of huntsman binaries requires no unusual assumptions, but instead is an expected phase for neutron star binaries with low-mass main sequence companions that have initial orbital periods above (but not much larger than) the bifurcation period.

Further study and discovery of  huntsman systems is a promising route to better understand the details of neutron star recycling. Neither confirmed huntsman has been fully timed, which would give a estimate of the current spindown luminosity and surface magnetic field. Better constraints on the neutron star and secondary masses, enabled by improved distance measurements and light curve modeling, would allow a determination of the efficiency of pulsar recycling at an intermediate stage in the process, and would also allow comparisons to the expected final pulsar--He white dwarf binaries. 

Owing to the 10s of Myr lifetime of the red bump phase, 
huntsman millisecond pulsars are likely to be intrinsically rare compared to typical spider pulsars, which have $\gtrsim$ Gyr lifetimes (e.g., \citealt{Chen13}). Nonetheless, the high luminosities of their secondaries should allow their discovery via optical follow-up at larger distances than for other spider pulsars. For example, known redbacks and black widows have median distances of $\sim 2$ kpc \citep{Strader19,Swihart2022}, compared to $\gtrsim 4$ kpc for the (albeit tiny) sample of huntsman pulsars. It seems likely that additional huntsman systems are present among the thousands of presently unassociated GeV $\gamma$-ray sources.

\section*{Acknowledgments}

We acknowledge the helpful comments from an anonymous referee, which improved the paper. We acknowledge support from NSF grant AST-2205550, NASA grants 80NSSC22K1583 and 80NSSC23K1350, and the Packard Foundation.

E.A. acknowledges support by NASA through the NASA Hubble Fellowship grant HST-HF2-51501.001-A awarded by the Space Telescope Science Institute, which is operated by the Association of Universities for Research in Astronomy, Inc., for NASA, under contract NAS5-26555.

Portions of this work performed at NRL were funded by NASA.

Based on observations obtained at the Southern Astrophysical Research (SOAR) telescope, which is a joint project of the Minist\'{e}rio da Ci\^{e}ncia, Tecnologia e Inova\c{c}\~{o}es (MCTI/LNA) do Brasil, the US National Science Foundation's NOIRLab, the University of North Carolina at Chapel Hill (UNC), and Michigan State University (MSU).

This work was based on observations obtained with XMM-Newton, an ESA science mission with instruments and contributions directly funded by ESA Member States and NASA.

We acknowledge the use of public data from the Swift data archive.

The Green Bank Observatory is a facility of the National Science Foundation operated under cooperative agreement by Associated Universities, Inc.

This research has made use of data and software provided by the High Energy Astrophysics Science Archive Research Center (HEASARC), which is a service of the Astrophysics Science Division at NASA/GSFC and the High Energy Astrophysics Division of the Smithsonian Astrophysical Observatory.

This work has made use of data from the European Space Agency (ESA) mission {\it Gaia} (\url{https://www.cosmos.esa.int/gaia}), processed by the {\it Gaia} Data Processing and Analysis Consortium (DPAC, \url{https://www.cosmos.esa.int/web/gaia/dpac/consortium}). Funding for the DPAC has been provided by national institutions, in particular the institutions participating in the {\it Gaia} Multilateral Agreement.

Based on observations obtained with the Samuel Oschin Telescope 48-inch and the 60-inch Telescope at the Palomar Observatory as part of the Zwicky Transient Facility project. ZTF is supported by the National Science Foundation under Grants No. AST-1440341 and AST-2034437 and a collaboration including current partners Caltech, IPAC, the Oskar Klein Center at
Stockholm University, the University of Maryland, University of California, Berkeley, the University of Wisconsin at Milwaukee, University of Warwick, Ruhr University, Cornell University, Northwestern University, and Drexel University. Operations are
conducted by COO, IPAC, and UW.

\clearpage
\bibliography{extracted}

\end{document}